# Mapping Relational Operations onto Hypergraph Model

Amani Naser Tahat [1, 2, 5], Maurice HT Ling[3, 4, 6]
[1]Department of Basic Science, Philadelphia University, Jordan
[2]Department of Physics, Hashemite University, Jordan
[3]School of Chemical and Life Sciences, Singapore Polytechnic, Singapore
[4]Department of Zoology, The University of Melbourne, Australia
[5]Amanitahat@yahoo.com; [6]mauriceling@acm.org

**Abstract**
The relational model is the most commonly used data model for storing large datasets, perhaps due to the simplicity of the tabular format which had revolutionized database management systems. However, many real world objects are recursive and associative in nature which makes storage in the relational model difficult. The hypergraph model is a generalization of a graph model, where each hypernode can be made up of other nodes or graphs and each hyperedge can be made up of one or more edges. It may address the recursive and associative limitations of relational model. However, the hypergraph model is non-tabular; thus, loses the simplicity of the relational model. In this study, we consider the means to convert a relational model into a hypergraph model in two layers. At the bottom layer, each relational tuple can be considered as a star graph centered where the primary key node is surrounded by non-primary key attributes. At the top layer, each tuple is a hypernode, and a relation is a set of hypernodes. We presented a reference implementation of relational operators (project, rename, select, inner join, natural join, left join, right join, outer join and Cartesian join) on a hypergraph model. Using a simple example, we demonstrate that a relation and relational operators can be implemented on this hypergraph model.

## 1. Introduction

A database can be seen as a structured collection of records that allows for proper storing, searching as well as retrieving of data. The conceptual organization of a database is known as the data model which describes the data, relationships between data elements and semantics along with data constraints. Data models can be categorized based on their underlying theoretical principles, such as the hierarchical model (Tsichritzis, 1976), the network model (CODASYL, 1974), and the relational model (Codd, 1970), as well as an emerging database model known as the object-oriented model (Arlow and Neustadt, 2001). Among them, the relational model (Codd, 1970), which is based on the set theory to construct data in terms of rows and columns and can be defined as a database that groups data by using common attributes found in the data set, is the most established and the most commonly used. Hence, this study shall focus on relational model.

The data model is crucial in terms of system design, functionality and maintainability (Deraman and Layzelll, 1995). It should reflect real world objects and their relationships to ensure durability.

A good data model serves and outlasts applications. Data models are domain and application specific. For example, data models which are useful for modeling inventories or financial records may well be different from data model which are important for modeling the domain of computer aided design applications or of genomic applications. Therefore, depending on the nature of the problem domain, new approaches might be needed.

Due to the simplicity of relational model in the operations of data storage and retrieval, relational databases have revolutionized database management systems. However, various shortcomings persist for use in data management needs for some domains. Firstly, poor representation of "real world" entities (considered as normalization) may lead to relations that do not keep up a correspondence to entities in "real world" (Reese, 2003). Secondly, the relational model lacks support for data-intensive and complex applications (Reese, 2003). Relational databases generally lack the ability to handle complex interrelationships of data such as images, and audio/video files and other digital files. Thirdly, relational databases require a homogeneous data structure which assumes both horizontal and vertical homogeneity (tabular form). However, most real-world objects are more complex. Thus, a homogeneous data structure is unnatural representation of real-world objects. Although many Relational Data Base Managements systems (RDBMSs) allow Binary Large Objects (BLOBs) (Shapiro and Miller, 1999), they are typically referenced to files, and as a result some advantages provided by DBMSs may be lost -- for instance the security. Furthermore, the inner structure of BLOBs cannot be accessed. Lastly, the relational model does not cater for semantic overloading. For example, the relational model only follows one construction for representing data as well as data relationships. Both concepts are presented by relation. Accordingly there will be no distinction between entity and relationship; no difference between different types of relations, then such semantic cannot be expressed (Chen, 1976).

Alternative models have been proposed to address the deficiencies of relational model, such as the graph model (Kunii, 1987), the graph-object oriented model (Gyssens et al., 1990), and the hypergraph model (Berge, 1973). A number of authors have found the hypergraph to be a useful means of modeling relational database designs (Beeri et al., 1981; Beeri et al., 1983; Chase, 1980; Fagin, 1983; Fagin et al., 1982; Yannakakis, 1981) as well as acting as an unifying data model for a wide variety of other data models (Angles and Gutierrez, 2008; Eschbach et al., 2006; Berge and Ray-Chaudhuri, 1972; Makinen, 1990). Sowa (1999) showed that any information can be represented by conceptual graphs.

Thus, we are motivated to use hypergraph model as a unified data model. In hypergraph, each node (known as a hypernode) can consist of one or more nodes or hypergraphs and each edge (known as a hyperedge) or link can consist of one or more edges.

As the relational model is widely taught and used, there appears to be a need to bridge between the relational model and the hypergraph model. New approaches, algorithms and theories for database query optimization have been developed that take advantage of advanced graph theoretic concepts. Hence, the relational model can be mapped onto a hypergraph model. Therefore, this study presents a reference implementation of mapping relational operators (Maier, 1983; Ullman, 1982) onto a hypergraph model.

## 2. Example of a Hypergraph Model

The hypergraph has played an important role in the recent research where the connection between databases and hypergraphs is straightforward and based on how to transform relational databases into graph databases in order to be able to reuse relational data already existent. The relational database schema can be considered as a set of attributes and a set of relations on those attributes. This can easily be represented by a hypergraph where the set of nodes in the hypergraph keep up a correspondence to the set of attributes in the database schema, along with each hyperedge corresponding to a set of attributes included in a relation in the database schema (Figure 1).

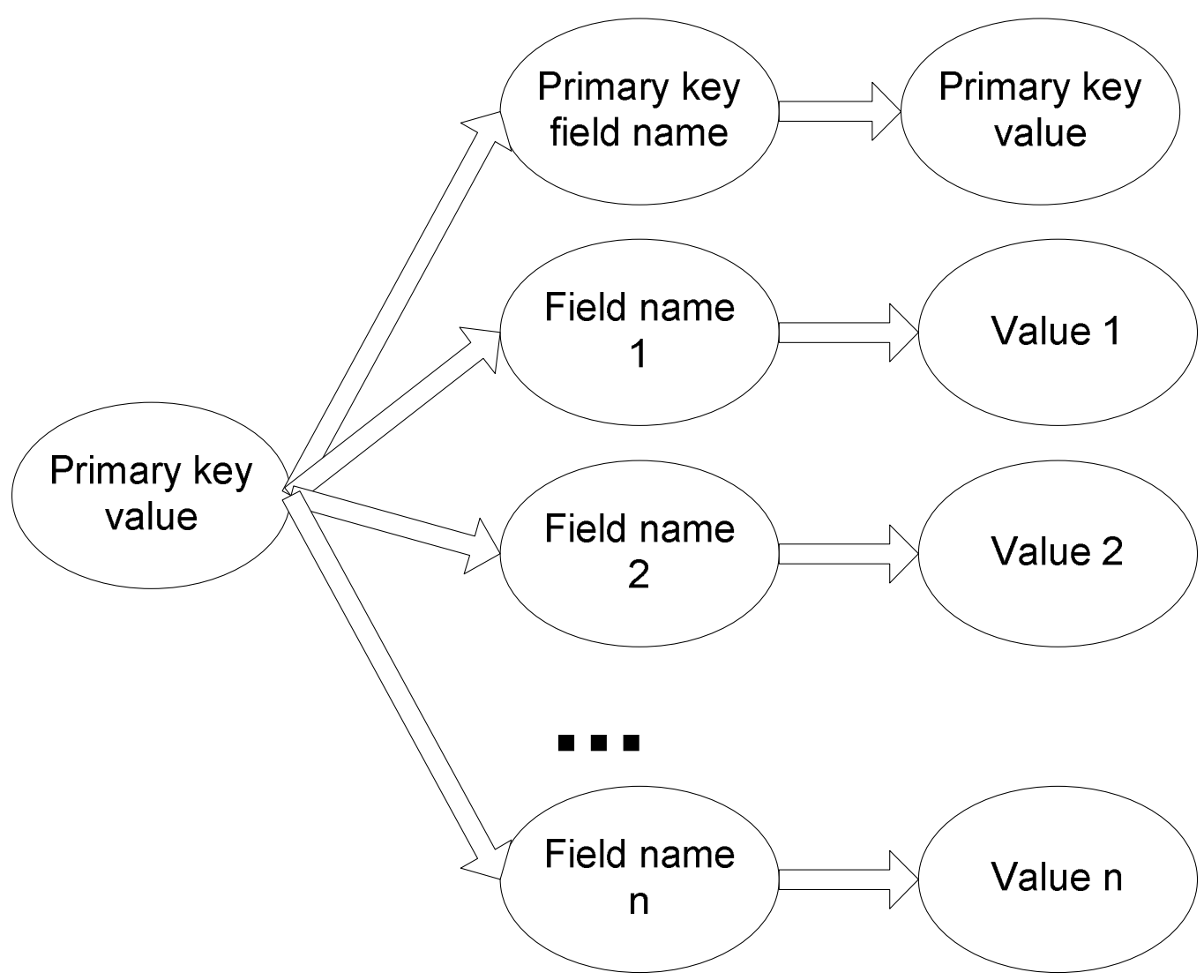

Figure 1: <u>The building block of the Graph.</u> The recognized definition of this graph can be written as:

```
<primary key value> :
{ <primary key field name> :
  <primary key value>,
  <Field name 1> : <Value 1>, …,
  <Field name n> : <Value n> }.
```

A combination of these graphs leads to a big graph close to most real world problems.

The issue of what we basing our implementation of a hypergraph-based DBMS is the similarity between a table in dictionary form and a graph in dictionary form. The data persistence mechanism used for the hypergraph is Python shelve module. Since "shelf" is a dictionary-like object, a graph can be made into a dictionary like:

```
graph [<origin node>] = {<destination node> : <attribute> +}
```

Similarly, a relational table can be constructed as a dictionary, such as:

```
table [<primary key value>] = {<primary key field name> : <primary key>, <field
                              name> : <data> +}
```

Our first task now is to map relational operations on a graph data model where the implementation of relational operations on graph (shelve) is that initial step:

Step 1: Construct 2 database tables using shelve in the format of:

```
table [<primary key value>] = {<primary key field name> : <primary key>,
<field name> : <data> +}
```

Step 2: Write functions to simulate the following relational operations: select, project, rename, inner join, left join, right join, outer join, inner join, Cartesian join (cross join), and natural join.

One may ask how to describe each edge in order to construct functions which will describe all edges for the graph model. It is quite an easy idea - each tuple (row) of a relational table is a star graph. Simply, if we have 2000 rows, we will have 2000 graphs. The central node is the primary key, other nodes are the fields and the data is on the edge as as a “star graph” shown in Figure 1.

Let’s use a simple library database to illustrate here. In a relational model, let a “books” table where ISBN is the primary key be defined as follows:

ISBN, Title, Publisher, First author, Catalog
-------------------------------------------------------------
9780596159818, Beautiful testing, O'Reilly, Tim Riley, 001
9781933988542, Open source SOA, Manning, Jeff Davis, 001
9780596516499, Natural language processing with Python, O'Reilly, Steven Bird, 001
9780521741033, Presentation skills for scientists, CUP, Edward Zanders, 002
9780751404624, E. coli, Blackie Academic, Chris Bell, 003

The catalog table will be:

Code, Description
--------------------------------------------------------------
001, computing
002, academic skills
003, biology

In graph model, the above tables will be:

```
Books = {'primary key': 'ISBN',
     '9780596159818': {'ISBN': '9780596159818',
                       'title': 'Beautiful testing',
                        'Publisher': "O'Reilly",
                        'first author': 'Tim Riley',
                        'catalog': '001'},
     '9781933988542': {'ISBN': '9781933988542',
                       'title': 'Open source SOA',
                        'publisher': 'Manning',
                        'first author': 'Jeff Davis',
                        'catalog': '001'},
     '9780596516499': {'ISBN': '9780596516499',
                       'title': 'Natural language processing Python',
                        'publisher': "O'Reilly",
```

```
                         'first author': 'Steven Bird',
                         'catalog': '001'},
     '9780521741033': {'ISBN': '9780521741033',
                         'title': 'Presentation skills for scientists',
                         'publisher': 'CUP',
                         'first author': 'Edward Zanders',
                         'catalog': '002'},
     '9780751404624': {'ISBN': '9780751404624',
                         'title': 'E. coli',
                         'publisher': 'Blackie Academic',
                         'first author': 'Chris Bell',
                         'catalog': '003'}
                 }
```

Hence, each book is a graph in the format of

```
<ISBN> ➔ <ISBN, title, publisher, first author>.
```

The catalog table can be:

```
Catalog = {'primary key': 'catalog',
     '001': {'catalog': '001', 'description': 'computing'},
     '002': {'catalog': '002', 'description': 'academic skills'},
     '003': {'catalog': '003', 'description': 'biology'}
     }
```

This means that we can define all the relational joins. For example, inner join can be

```
def inner_join(<left table>, <right table>, <joining field>)
```

The set of eight relational operations (project, select, inner join, left join, right join, outer join, inner join, Cartesian product) can then form the reference idea to implement more complicated operations.

## 3. Reference Implementation of Relational Operators on Hypergraph Model

This reference implementation employs the one-table, one-file approach where each table is persisted in a shelve file. Therefore, adding new tables we only need to call the function `shelve.open(<table name>)` and we can write any data to it, as illustrated in the following example:

```
newtable = shelve.open('new_table')
```

Each data record (tuple) will be a dictionary identified by the primary key value in the format of `newtable[ <primary key> ] = { <field name #1>: <value #1>, … }`. For example, the following will add data into the books and catalog data tables:

```
Books['9780596159819'] = {'ISBN': '9780596159819',
                          'title': 'New book',
                          'publisher': 'Amani',
                          'first author': 'Valeriy',
                          'catalog': '004'}

Catalog['005'] = {'catalog': '005', 'description': 'News'}
```

Deleting records can then be performed using dictionary operations.

Select, project and rename operators can be implemented as:

```
def rename(dic, old_key, new_key):
    """Rename operator.

    @param dic: table (tuple) represented as graph
    @type dic: 2-level dictionary
    @param old_key: original field (attribute) name
    @type: string
    @param new_key: new field (attribute) name to be renamed to
    @type new_key: string

    @return: field name renamed table represented as graph
    """
    for k in dic:
        for kk in dic[k]:
            if(kk == old_key):
                dic[k][new_key] = dic[k][old_key]
                del(dic[k][old_key])
    return dic

def select(db, where=''):
    """Select (restriction) operator.

    @param db: table(tuple) represented as graph
    @type db: 2-level dictionary
    @param where: selection condition in the format of <field name>=<condition>

    @return: selected (restricted) table represented as graph
    """
    table = []
    if (len(where) > 0 ):
        where = where.split('=')
        where[0] = where[0].strip()
        where[1] = where[1].strip()
    ret = {}
    for k in db:
        try:
            field = ''
            if (len(where) == 2):
                field = db[k][where[0]]
            if((len(where) == 2 and field == where[1]) or len(where) == 0):
                ret[k] = db[k]
```

```
        except KeyError: pass
    return ret

def project(columns, db):
    """Projection operator.

    @param columns: comma-delimited list of fields (attributes) to project
    @type columns: string
    @param db: table(tuple) represented as graph
    @type db: 2-level dictionary

    @return: projected table represented as graph
    """
    columns = [x.strip() for x in columns.split(',')]
    ret = {}
    for k in db:
        if(columns[0] == '*'):
            ret[k] = db[k]
        else:
            ret[k] = {}
            for kk in columns:
                try:
                    ret[k][kk] = db[k][kk]
                except KeyError: pass
    return ret
```

Given two tables designated as the ‘right’ and ‘left’ tables of a join operation and the ‘keys’ being the joining field, various join operations (left, right, inner, outer, Cartesian, and natural) can be implemented as:

```
def left_join(left, right, key=None):
    """
    Left join operator.

    @param left: table (tuple) at the left side of the join
    @type left: dictionary
    @param right: table (tuple) at the right side of the join
    @type right: dictionary
    @param key: name of joining field (attribute)
    @type key: string

    @return: joined table represented as graph
    """
    if not key:
        return {}
    ret = {}
    for k in left:
            ret[k] = left[k]
            if(right.has_key(left[k][key])):
                ret[k][key] = right[left[k][key]]
            ret[k] = flatten(ret[k])
    return ret
```

```
def inner_join(left, right, key=None):
    """
    Inner join operator

    @param left: table (tuple) at the left side of the join
    @type left: dictionary
    @param right: table (tuple) at the right side of the join
    @type right: dictionary
    @param key: name of joining field (attribute)
    @type key: string

    @return: joined table represented as graph
    """
    if not key:
        return {}
    ret = {}
    for k in left:
        if(right.has_key(left[k][key])):
            if(len(right[left[k][key]]) > 0):
                ret[k] = left[k]
                ret[k][key] = right[left[k][key]]
                ret[k] = flatten(ret[k])
    return ret

def right_join(left, right, key=None):
    """
    Right join operator

    @param left: table (tuple) at the left side of the join
    @type left: dictionary
    @param right: table (tuple) at the right side of the join
    @type right: dictionary
    @param key: name of joining field (attribute)
    @type key: string

    @return: joined table represented as graph
    """
    ret = inner_join(left, right, key)
    for k in right:
        found = 0
        empty ={}
        for kk in left:
            if(left[kk][key] == k):
                found = 1
        if found == 0:
            left_keys = left.keys()
            left_row = left[left_keys[0]]
            for kk in left_row:
                if kk == key:
                    empty[kk] = right[k]
                else:
                    empty[kk] = ''
            ret[k] = empty
            ret[k] = flatten(ret[k])
```

```
    return ret

def outer_join(left, right, key=None):
    """
    Outer join operator

    @param left: table (tuple) at the left side of the join
    @type left: dictionary
    @param right: table (tuple) at the right side of the join
    @type right: dictionary
    @param key: name of joining field (attribute)
    @type key: string

    @return: joined table represented as graph
    """
    ret = left_join(left, right, key)
    for k in right:
        found = 0
        empty ={}
        for kk in left:
            if(left[kk][key] == k):
                found = 1
        if found == 0:
            left_keys = left.keys()
            left_row = left[left_keys[0]]
            for kk in left_row:
                if kk == key:
                    empty[kk] = right[k]
                else:
                    empty[kk] = ''
            ret[k] = empty
            ret[k] = flatten(ret[k])
    return ret

def cartesian(left, right, keys=None):
    """
    Cartesian (cross) join operator

    @param left: table (tuple) at the left side of the join
    @type left: dictionary
    @param right: table (tuple) at the right side of the join
    @type right: dictionary
    @param key: name of joining field (attribute)
    @type key: string

    @return: joined table represented as graph
    """
    if not keys:
        return {}
    ret = {}
    for left_key in left:
        for right_key in right:
            new_key = left_key + '_' + right_key
            ret[new_key] = left[left_key]
```

```
            ret[new_key][keys] = right[right_key]

            ret[new_key] = flatten(ret[new_key])
    for k in right:
        found = 0
        empty = {}
        for kk in left:
            if(left[kk][keys] == k):
                found = 1
        if found == 0:
            for kk in left:
                key = k + '_' + kk
                ret[key] = left[kk]
                ret[key][keys] = right[k]

                ret[key] = flatten(ret[key])
    return ret

def natural_join(left, right):
    """
    Natural join operator

    @param left: table (tuple) at the left side of the join
    @type left: dictionary
    @param right: table (tuple) at the right side of the join
    @type right: dictionary

    @return: joined table represented as graph
    """
    for k in left:
        keys1 = left[k].keys()
        for kk in right:
            keys2 = right[kk].keys()
            break
        break
    key = ""
    for k in keys1:
        for kk in keys2:
            if k == kk:
                key = k
                break
        if key != '':
            break
    if key == '':
        return {}
    ret = {}
    for k in left:
        if(right.has_key(left[k][key])):
            if(len(right[left[k][key]]) > 0):
                ret[k] = left[k]
                ret[k][key] = right[left[k][key]]
                ret[k] = flatten(ret[k])
    return ret
```

Joining 2 tables will result in a nested dictionary. Taking left join as an example, data from the right table (dictionary) will be inserted in the joining field of the left table. Hence, we implemented a helper function (modified from Terry Jones, 18 February 2009 post on comp.lang.python – flatten a dict), flatten, to remove the nesting.

```
def flatten(d, prefix=None, sep='.'):
    """
    Flattening of a 3-level dictionary structure after join operations into a
    2-level dictionary structure. Modified from Terry Jones, 18 February 2009
    post on comp.lang.python – flatten a dict.

    @param d: dictionary to flatten
    @type d: 2-level or 3-level dictionary of dictionary
    @param prefix: prefix in flattened keys, default=None
    @type prefix: string
    @param sep: separator for flattened keys, default='.'
    @type sep: string

    @return: 2-level flattened dictionary
    """
    result = {}
    if prefix is None:
        prefix = ''
    for k, v in d.iteritems():
        if prefix:
          key = prefix + sep + k
      else:
          key = k
        if isinstance(v, dict):
           if v:
                result.update(flatten(v,key))
           else:
               result[key] = None
        else:
            result[key] = v
    return result
```

## 4. Testing the reference implementation

The following is a test command of our implementation of relational operators.

```
import sys
import os
import shelve
import unittest

Books = {
         '9780596159818': {'ISBN': '9780596159818',
                           'title': 'Beautiful testing',
                           'publisher': "O'Reilly",
                           'first author': 'Tim Riley',
```

```
                          'catalog': '001'},
         '9781933988542': {'ISBN': '9781933988542',
                          'title': 'Open source SOA',
                          'publisher': 'Manning',
                          'first author': 'Jeff Davis',
                          'catalog': '001'},
         '9780596516499': {'ISBN': '9780596516499',
                          'title': 'Natural language processing Python',
                          'publisher': "O'Reilly",
                          'first author': 'Steven Bird',
                          'catalog': '001'},
         '9780521741033': {'ISBN': '9780521741033',
                          'title': 'Presentation skills for scientists',
                          'publisher': 'CUP',
                          'first author': 'Edward Zanders',
                          'catalog': '002'},
         '9780751404624': {'ISBN': '9780751404624',
                          'title': 'E. coli',
                          'publisher': 'Blackie Academic',
                          'first author': 'Chris Bell',
                          'catalog': '003'}
        }

Catalog = {
           '001': {'catalog': '001', 'description': 'computing'},
           '002': {'catalog': '002', 'description': 'academic skills'},
           '003': {'catalog': '003', 'description': 'biology'}
          }

class testRelational(unittest.TestCase):

    def teardown(self):
        books.close()
        catalog.close()

    def testShelveTableIntegrity_1(self):
        """Check that datatables are identical to test data - equality
        check"""
        self.assertTrue(books == Books)
        self.assertTrue(catalog == Catalog)

    def testShelveTableIntegrity_2(self):
        """Check that datatables are identical to test data - length check"""
        self.assertEqual(len(books), len(Books))
        self.assertEqual(len(catalog), len(Catalog))

    def testSelect_Full(self):
        """Selecting entire table"""
        self.assertTrue(books, select(books))
        self.assertTrue(catalog, select(catalog))

    def testSelect_Single(self):
        """Select publisher=O'Reilly on books table"""
        result = {'9780596159818': {'ISBN': '9780596159818',
```

```
                         'title': 'Beautiful testing',
                         'publisher': "O'Reilly",
                         'first author': 'Tim Riley',
                         'catalog': '001'},
                  '9780596516499': {'ISBN': '9780596516499',
                         'title': 'Natural language processing Python',
                         'publisher': "O'Reilly",
                         'first author': 'Steven Bird',
                         'catalog': '001'}
            }
    self.assertTrue(result == select(books, "publisher=O'Reilly"))

def testSelect_Double(self):
    """Select publisher=O'Reilly and first author=Steven Bird on books
    table"""
    result = {'9780596516499': {'ISBN': '9780596516499',
                         'title': 'Natural language processing Python',
                         'publisher': "O'Reilly",
                         'first author': 'Steven Bird',
                         'catalog': '001'}
            }
    self.assertTrue(result == select(select(books, "publisher=O'Reilly"),
                                  "first author=Steven Bird"))

def testProject_Full(self):
    """Project everything after selecting publisher=O'Reilly on books
    table"""
    result = {'9780596159818': {'ISBN': '9780596159818',
                         'title': 'Beautiful testing',
                         'publisher': "O'Reilly",
                         'first author': 'Tim Riley',
                         'catalog':'001'},
              '9780596516499': {'ISBN': '9780596516499',
                         'title': 'Natural language processing Python',
                         'publisher': "O'Reilly",
                         'first author': 'Steven Bird',
                         'catalog': '001'}
             }
    temp = select(books, "publisher=O'Reilly")
    self.assertTrue(result == project('*',temp))

def testProject_TwoFields(self):
    """
    Project title and catalog after selecting publisher=O'Reilly on books
    table"""
    result = {'9780596159818': {'title': 'Beautiful testing',
                                'catalog': '001'},
              '9780596516499': {'title':
                                  'Natural language processing Python',
                               'catalog': '001'}
             }
    temp = select(books, "publisher=O'Reilly")
    self.assertTrue(result == project('title, catalog',temp))
```

```
def testRename_1(self):
    """Rename description field to category field on catalog table"""
    result = {
           '001': {'catalog': '001', 'category': 'computing'},
           '002': {'catalog': '002', 'category': 'academic skills'},
           '003': {'catalog': '003', 'category': 'biology'}
             }
    self.assertTrue(result == rename(project("*",catalog),
                                 'description', 'category'))

def testRename_2(self):
    """
    Rename description field to category field and catalog field to code
    field on catalog table"""
    result = {
           '001': {'code': '001', 'category': 'computing'},
           '002': {'code': '002', 'category': 'academic skills'},
           '003': {'code': '003', 'category': 'biology'}
             }
    self.assertTrue(result == rename(rename(project("*",catalog),
                                        'description', 'category'),
                                 'catalog', 'code'))

def testInnerJoin_Correct(self):
    """Inner join of books table and catalog table on catalog=catalog.
    This is equivalent to natural join"""
    self.assertEqual(inner_join(books, catalog, 'catalog'),
                     natural_join(books, catalog))

def testNaturalJoin(self):
    """Natural join of books table and catalog table, which will join on
    The catalog field of both tables"""
    result = {'9780751404624': {'publisher': 'Blackie Academic',
                          'catalog.catalog': '003',
                          'catalog.description': 'biology',
                          'first author': 'Chris Bell',
                          'ISBN': '9780751404624',
                          'title': 'E. coli'},
              '9780596159818': {'publisher': "O'Reilly",
                          'catalog.catalog': '001',
                          'catalog.description': 'computing',
                          'first author': 'Tim Riley',
                          'ISBN': '9780596159818',
                         'title': 'Beautiful testing'},
              '9781933988542': {'publisher': 'Manning',
                          'catalog.catalog': '001',
                          'catalog.description': 'computing',
                          'first author': 'Jeff Davis',
                          'ISBN': '9781933988542',
                          'title': 'Open source SOA'},
              '9780521741033': {'publisher': 'CUP',
                          'catalog.catalog': '002',
                          'catalog.description': 'academic skills',
                          'first author': 'Edward Zanders',
```

```
                             'ISBN': '9780521741033',
                             'title': 'Presentation skills for scientists'},
              '9780596516499': {'publisher': "O'Reilly",
                             'catalog.catalog': '001',
                             'catalog.description': 'computing',
                             'first author': 'Steven Bird',
                             'ISBN': '9780596516499',
                             'title': 'Natural language processing Python'}
             }
    self.assertTrue(result == natural_join(books, catalog))

def testProjectAfterInnerJoin_Select(self):
    """Combining inner join, select, and project"""
    result = {'9780596159818': {'catalog.description': 'computing',
                             'title': 'Beautiful testing'},
              '9781933988542': {'catalog.description': 'computing',
                             'title': 'Open source SOA'},
              '9780596516499': {'catalog.description': 'computing',
                             'title': 'Natural language processing Python'}
             }
    temp = inner_join(books, catalog, "catalog")
    temp = select(temp, "catalog.catalog=001")
    temp = project("title, catalog.description",temp)
    self.assertTrue(result == temp)

def testLeftJoin(self):
    """Left join books and catalog"""
    result = {'9780751404624': {'publisher': 'Blackie Academic',
                         'catalog.catalog': '003',
                         'catalog.description': 'biology',
                         'first author': 'Chris Bell',
                         'ISBN': '9780751404624',
                         'title': 'E. coli'},
              '9780596159818': {'publisher': "O'Reilly",
                         'catalog.catalog': '001',
                         'catalog.description': 'computing',
                         'first author': 'Tim Riley',
                         'ISBN': '9780596159818',
                         'title': 'Beautiful testing'},
              '9781933988542': {'publisher': 'Manning',
                         'catalog.catalog': '001',
                         'catalog.description': 'computing',
                         'first author': 'Jeff Davis',
                         'ISBN': '9781933988542',
                         'title': 'Open source SOA'},
              '9780521741033': {'publisher': 'CUP',
                         'catalog.catalog': '002',
                         'catalog.description':
                         'academic skills',
                         'first author': 'Edward Zanders',
                         'ISBN': '9780521741033',
                         'title': 'Presentation skills for scientists'},
              '9780596516499': {'publisher': "O'Reilly",
                         'catalog.catalog': '001',
```

```
                        'catalog.description': 'computing',
                        'first author': 'Steven Bird',
                        'ISBN': '9780596516499',
                        'title': 'Natural language processing Python'}
             }
    self.assertTrue(result == left_join(books, catalog,'catalog'))

def testRightJoin(self):
    """Right join books and catalog"""
    result = {'9780751404624': {'publisher': 'Blackie Academic',
                        'catalog.catalog': '003',
                        'catalog.description': 'biology',
                        'first author': 'Chris Bell',
                        'ISBN': '9780751404624',
                        'title': 'E. coli'},
              '9780596159818': {'publisher': "O'Reilly",
                        'catalog.catalog': '001',
                        'catalog.description': 'computing',
                        'first author': 'Tim Riley',
                        'ISBN': '9780596159818',
                        'title': 'Beautiful testing'},
              '9781933988542': {'publisher': 'Manning',
                        'catalog.catalog': '001',
                        'catalog.description': 'computing',
                        'first author': 'Jeff Davis',
                        'ISBN': '9781933988542',
                        'title': 'Open source SOA'},
              '9780521741033': {'publisher': 'CUP',
                        'catalog.catalog': '002',
                        'catalog.description': 'academic skills',
                        'first author': 'Edward Zanders',
                        'ISBN': '9780521741033',
                        'title': 'Presentation skills for scientists'},
              '9780596516499': {'publisher': "O'Reilly",
                        'catalog.catalog': '001',
                        'catalog.description': 'computing',
                        'first author': 'Steven Bird',
                        'ISBN': '9780596516499',
                        'title': 'Natural language processing Python'}
             }
    self.assertTrue(result == right_join(books, catalog,'catalog'))

def testOuterJoin(self):
    """Outer join books and catalog"""
    result = {'9780751404624': {'publisher': 'Blackie Academic',
                        'catalog.catalog': '003',
                        'catalog.description': 'biology',
                        'first author': 'Chris Bell',
                        'ISBN': '9780751404624',
                        'title': 'E. coli'},
              '9780596159818': {'publisher': "O'Reilly",
                        'catalog.catalog': '001',
                        'catalog.description': 'computing',
                        'first author': 'Tim Riley',
```

```
                    'ISBN': '9780596159818',
                    'title': 'Beautiful testing'},
          '9781933988542': {'publisher': 'Manning',
                    'catalog.catalog': '001',
                    'catalog.description': 'computing',
                    'first author': 'Jeff Davis',
                    'ISBN': '9781933988542',
                    'title': 'Open source SOA'},
          '9780521741033': {'publisher': 'CUP',
                    'catalog.catalog': '002',
                    'catalog.description': 'academic skills',
                    'first author': 'Edward Zanders',
                    'ISBN': '9780521741033',
                    'title': 'Presentation skills for scientists'},
          '9780596516499': {'publisher': "O'Reilly",
                    'catalog.catalog': '001',
                    'catalog.description': 'computing',
                    'first author': 'Steven Bird',
                    'ISBN': '9780596516499',
                    'title': 'Natural language processing Python'}
         }
    self.assertTrue(result == outer_join(books, catalog,'catalog'))

def testCartesianJoin_Full(self):
    """Full cartesian join, also known as cross join"""
    result = {'9780596516499_003': {'publisher': "O'Reilly",
                    'catalog.catalog': '003',
                    'catalog.description': 'biology',
                    'first author': 'Steven Bird',
                    'ISBN': '9780596516499', 'title':
                    'Natural language processing Python'},
          '9780596516499_002': {'publisher': "O'Reilly",
                    'catalog.catalog': '002',
                    'catalog.description': 'academic skills',
                    'first author': 'Steven Bird',
                    'ISBN': '9780596516499',
                    'title': 'Natural language processing Python'},
          '9780596516499_001': {'publisher': "O'Reilly",
                    'catalog.catalog': '001',
                    'catalog.description': 'computing',
                    'first author': 'Steven Bird',
                    'ISBN': '9780596516499',
                    'title': 'Natural language processing Python'},

          '9781933988542_001': {'publisher': 'Manning',
                    'catalog.catalog': '001',
                    'catalog.description': 'computing',
                    'first author': 'Jeff Davis',
                    'ISBN': '9781933988542',
                    'title': 'Open source SOA'},
          '9781933988542_002': {'publisher': 'Manning',
                    'catalog.catalog': '002',
                    'catalog.description': 'academic skills',
                    'first author': 'Jeff Davis',
```

```
          'ISBN': '9781933988542',
          'title': 'Open source SOA'},
'9781933988542_003': {'publisher': 'Manning',
          'catalog.catalog': '003',
          'catalog.description': 'biology',
          'first author': 'Jeff Davis',
          'ISBN': '9781933988542',
          'title': 'Open source SOA'},

'9780521741033_002': {'publisher': 'CUP',
          'catalog.catalog': '002',
          'catalog.description': 'academic skills',
          'first author': 'Edward Zanders',
          'ISBN': '9780521741033',
          'title': 'Presentation skills for scientists'},
'9780521741033_003': {'publisher': 'CUP',
          'catalog.catalog': '003',
          'catalog.description': 'biology',
          'first author': 'Edward Zanders',
          'ISBN': '9780521741033',
          'title': 'Presentation skills for scientists'},
'9780521741033_001': {'publisher': 'CUP',
          'catalog.catalog': '001',
          'catalog.description': 'computing',
          'first author': 'Edward Zanders',
          'ISBN': '9780521741033',
          'title': 'Presentation skills for scientists'},

'9780751404624_001': {'publisher': 'Blackie Academic',
          'catalog.catalog': '001',
          'catalog.description': 'computing',
          'first author': 'Chris Bell',
          'ISBN': '9780751404624',
          'title': 'E. coli'},
'9780751404624_003': {'publisher': 'Blackie Academic',
          'catalog.catalog': '003',
          'catalog.description': 'biology',
          'first author': 'Chris Bell',
          'ISBN': '9780751404624',
          'title': 'E. coli'},
'9780751404624_002': {'publisher': 'Blackie Academic',
          'catalog.catalog': '002',
          'catalog.description': 'academic skills',
          'first author': 'Chris Bell',
          'ISBN': '9780751404624',
          'title': 'E. coli'},

'9780596159818_002': {'publisher': "O'Reilly",
          'catalog.catalog': '002',
          'catalog.description': 'academic skills',
          'first author': 'Tim Riley',
          'ISBN': '9780596159818',
          'title': 'Beautiful testing'},
'9780596159818_003': {'publisher': "O'Reilly",
```

```
                              'catalog.catalog': '003',
                              'catalog.description': 'biology',
                              'first author': 'Tim Riley',
                              'ISBN': '9780596159818',
                              'title': 'Beautiful testing'},
                    '9780596159818_001': {'publisher': "O'Reilly",
                              'catalog.catalog': '001',
                              'catalog.description': 'computing',
                              'first author': 'Tim Riley',
                              'ISBN': '9780596159818',
                              'title': 'Beautiful testing'}
                   }
        self.assertTrue(result, cartesian(books, catalog, 'catalog'))

    def testCartesianJoin_Select(self):
        """Select for 'first author=Chris Bell' after cartesian join."""
        result = {'9780751404624_001': {'publisher': 'Blackie Academic',
                              'catalog.catalog': '001',
                              'catalog.description': 'computing',
                              'first author': 'Chris Bell',
                              'ISBN': '9780751404624',
                              'title': 'E. coli'},
                    '9780751404624_003': {'publisher': 'Blackie Academic',
                              'catalog.catalog': '003',
                              'catalog.description': 'biology',
                              'first author': 'Chris Bell',
                              'ISBN': '9780751404624',
                              'title': 'E. coli'},
                    '9780751404624_002': {'publisher': 'Blackie Academic',
                              'catalog.catalog': '002',
                              'catalog.description': 'academic skills',
                              'first author': 'Chris Bell',
                              'ISBN': '9780751404624',
                              'title': 'E. coli'}
                   }
        temp = cartesian(books, catalog, 'catalog')
        temp = select(temp, 'first author=Chris Bell')
        self.assertTrue(result, temp)

if __name__ == '__main__':
    books = shelve.open('books.db')
    catalog = shelve.open('catalog.db')
    for key in books: del books[key]
    for key in catalog: del catalog[key]
    for key in Books: books[key] = Books[key]
    for key in Catalog: catalog[key] = Catalog[key]
    unittest.main()
```

## 5. Conclusion

This study attempts to unify the most commonly used model (relational database model) into a graph-based model. The path of least dependency has been implemented by using the shelve module in Python standard library. Relational operators; such as select, project, rename, inner join, left join, right join, outer join, natural join and Cartesian product; have been implemented to operate on dictionary-like databases based on a hypergraph model. The analogy between tables and graphs has been shown to define a process for converting databases organized according to the relational model to a database that is based on a hypergraph model.